\documentclass[CRPHYS,Unicode,manuscript]{cedram}

\usepackage{aas_macros}

\usepackage[normalem]{ulem} 

\title{Prospects for the characterization of habitable planets}

\author{\firstname{Stephane} \lastname{Mazevet}\IsCorresp}
\address{Observatoire de la C\^ote d'Azur, Universit\'e C\^ote d'Azur, CNRS, 96 boulevard de l'observatoire, F06304 Nice cedex 4, France}
\email[S. Mazevet]{stephane.mazevet@oca.eu}
\author{\firstname{Antonin} \lastname{Affholder}}
\address{Institut de Biologie de l’\'Ecole Normale Supérieure, ENS, Universit\'e Paris Sciences et Lettres, Paris, France}
\author{\firstname{Boris} \lastname{Sauterey}}
\address{Institut de Biologie de l’\'Ecole Normale Supérieure, ENS, Universit\'e Paris Sciences et Lettres, Paris, France}
  \author{\firstname{Alex} \lastname{Bixel}}
  \author{\firstname{Daniel} \lastname{Apai}}
  \address{Department of Astronomy, The University of Arizona, Tucson, AZ 85721, USA, and\\
    Lunar and Planetary Laboratory, The University of Arizona, Tucson, AZ 85721, USA}
  \author{\firstname{Regis} \lastname{Ferriere}}
 \address{Institut de Biologie de l’\'Ecole Normale Supérieure, ENS, Universit\'e Paris Sciences et Lettres, Paris, France\\
  Department of Ecology and Evolutionary Biology, University of Arizona, Tucson, USA, and\\
  International Research Laboratory for Interdisciplinary Global Environmental Studies (iGLOBES), CNRS, ENS, Université Paris Sciences et Lettres, University of Arizona, Tucson, USA}


\begin{abstract} 
  With thousands of exoplanets now identified, the characterization of habitable planets and the potential identification of inhabited ones
  is a major challenge for the coming decades. We review the current working definition of habitable planets, the upcoming observational prospects for
  their characterization and present an innovative approach to assess habitability and inhabitation. This integrated method couples for the first time
  the atmosphere and the interior modeling with the biological activity based on ecosystem modeling. We review here the first applications of the method
  to asses the likelihood and impact of methanogenesis for Enceladus, primitive Earth, and primitive Mars. Informed by these applications for solar system situations where
  habitability and inhabitation is questionned, we show how the method can be used to inform the design of future space observatories by considering habitability and inhabitation of
  Earth-like exoplanets around sun-like stars.  
\end{abstract}

\begin{document}

\maketitle

\selectlanguage{english}


A little more than 25 years after the discovery of the first exoplanet\cite{mayor1995}, we have now identified more than 5,000 objects ranging in size
from mercury-like objects to ones that have several times the mass of Jupiter with orbital periods spanning from less than a day to several decades. This broad diversity
of planets when considering their size, their composition or their orbital properties largely encompasses what is found in the solar system. Understanding this diversity
as well as the nature of the objects detected is currently a major challenge in planetary science.
It is generating an intense activity to design new observational strategies for either ground and space observatories as, at present, the majority of planets are detected indirectly,
by quantifying their influence on their host stars. Using radial velocities or transit methods\cite{perryman2018}, the parameters of a given planet are poorly constrained, and are
restricted to the mass, the radius and a few orbital parameters. The strongest constraints are obtained when these methods can be combined.
Atmospheric characterization, that is the basis of our knowledge of the planets of the solar system, is currently very limited for exoplanets.

Atmospheric characterization of exoplanets has only been obtained in an handful of cases. This corresponds to either giant planets with shorts periods that can be characterized using
transit spectroscopy\cite{sing2016} or, conversely, far-out planets in the giant-brown-dwarf continuum that can be accessed using direct imaging\cite{charnay2019,sphere}.
The upcoming generation of instruments from either ground or space such as, for example, the SPHERE and GRAVITY upgrades\cite{sphere+,gravity2019}, the European Extremely Large
Telescope (E-ELT), the James Web Space Telescope (JWST)\cite{jwst} and the upcoming ARIEL mission\cite{ariel2018} from ESA, aim at improving this situation by characterizing the
atmosphere of smaller objects and at varying distance from their host star.
While addressing various questions along the way, such as the atmospheric composition of giant planets, the nature of ocean worlds, of sub-neptunes and super-earths, the ultimate and
long term goals of the next generation of instruments is to enable us to quantify the habitability of Earth-like planets and eventually identify inhabited ones.  In this paper, we review
the current status regarding the prospects for the characterization of habitable exoplanets and present an innovative framework to asses habitability and inhabitation on quantitative
grounds so as to inform the design of the next generation of space observatories.

\section{Current status on habitable planets}
\subsection{Definition of the habitable zone}

Brought forward by H. Shapley more than seventy years ago\cite{Catling2017}, the concept of the \emph{habitable zone} rests on the idea that there exists a region around each star where
liquid water can be maintained at the surface of an Earth-like planet. The connection between habitability and liquid water is obviously related to the fact that life as we know it
requires liquid water to thrive and develop. This definition of habitability relates to life that resembles what we find on Earth and relates to one of its basic constituent, water. It
secondly conveys the idea that life is a surface event that, in turn, may potentially modify the atmospheric composition of the planet if present. Planets with liquid water in the
subsurface and protected by an ice layer can also be described as habitable. They may exist outside the habitable zone that is now defined by the possibility of surface liquid water but
will not be considered here as they are poor candidates for remote sensing based on atmospheric characterization.  

The modern estimation of the boundaries of the habitable zone is obtained by using climate modeling for an atmosphere made of CO$_2$, H$_2$O, and N$_2$\cite{kasting1993}.
The inner edge is defined as the distance at which a planet develops a wet stratosphere and loses its water through
photodissociation and the escape of hydrogen to space\cite{kasting1993,kopparapu2013}. It is the so-called water-loss limit where the planet enters a wet greenhouse runaway.
It occurs before the hard greenhouse runaway limit where the atmosphere absorbs more energy than it can radiate back to space. In this case, the surface temperature
rapidly increases beyond one thousand degrees and well above what can be deemed habitable. For todays solar system, these two limits are estimated to be at 0.97 AU
and 0.99 AU using a 1-D climate modeling with a fully saturated troposphere and a cloud free atmosphere\cite{kopparapu2013}. Numerous 3D climate modeling studies
have refined these values (see below). 

The outer edge of the habitable zone is set by the maximum greenhouse effect that can be obtained by continuously adding CO$_2$ in a H$_2$O-CO$_2$ atmosphere.
This can occur when a planet experiences an active carbonate-silicate cycle that can lead to a continuous build-up of CO$_2$. Beyond a certain amount of CO$_2$ in
the atmosphere, which reaches around 7-8 bars, CO$_2$ condensates and Rayleigh scattering takes over the greenhouse effect. In this case, the planet experiences
cooling instead of warming. 1-D modeling sets the outer limit of the habitable zone at 1.67 AU for todays sun\cite{kopparapu2013}. Formaly, the habitable zone could
be extended outward assuming additional greenhouse gases, such as NH$_3$ and CH$_4 $, in the atmosphere but most gases will not significantly move the outer limit. The exception is hydrogen that can
seriously enhance the greenhouse effect when considering dense atmospheres\cite{stevenson1999,pierrehumbert2011,ramirez2014,ramirez2017}.

Similar 1-D climate calculations are used to estimate the boundaries of the habitable zone for other type of stars. These calculations take into account the fact that the luminosity
of the star decreases with its mass while its emission spectrum shifts to the red and to larger wavelengths. As the stellar mass decreases, the habitable zone gets closer to the
host star and enters a region where the planets are tidally locked (i.e., the planetary rotation and orbit are synchronized). This inner and outer limits obtained using climate
calculations set what is called the conservative habitable zone. This estimate has been refined by considering additional greenhouses gases as mentionned above and using complex 3-D climate
models such as 3D general circulation models (GCM). The latter take into account various effects such as cloud formation, atmospheric circulation, rotation rate. These 3-D estimates of
the habitable zone are within 5 to 7 \% of the 1-D ones for fast rotating planets like the Earth in the habitable zone of (F,G,K) stars. They predict much larger differences
for tidally locked and slowly rotating planets that lay in the habitable zone of (K,M) stars. In this situation, 3-D modeling predicts the inner edge of the habitable zone much closer
to their host star.      

This estimate of the conservative habitable zone, introduced by Kasting {\sl et al.}\cite{kasting1993}, updated by Kopparapu {\sl et al.}\cite{kopparapu2013,kopparapu2014} and augmented
by 3-D modeling\cite{leconte2013} is directly related to the modeling of the long term climate evolution of the inner planets of the solar system\cite{kasting1993,Catling2017}. As it might
be perceived as too restrictive and prone to modeling shortcomings, it is often associated with another empirical estimate directly tied to observables associated to Venus and Mars.
In this case, the optimistic inner edge of habitable zone rests on the observation that the age of Venus young cratering are estimated to be at about 0.7 Gyrs.
This provides a lower limit to its water loss by either wet or greenhouse run away. Taking into account the sun luminosity at that time, this sets the boundary for
the greenhouse runaway at 0.75AU. On the other side of the habitable zone, the surface of Mars suggests that water may have been present 3.8 Gyrs ago. Considering that the solar luminosity
was 75\% of its current value, this sets the outer boundary of the optimistic habitable zone at 1.75AU, past the estimate obtained using climate modeling, 1.69AU, suggesting that other
greenhouse gases that CO$_2$ and H$_2$O were at play if primitive Mars was habitable\cite{ramirez2014}.

\subsection{The nature of the planets identified as potentially habitable}
The previous section underlines that the habitable zone is not a planetary property but a region around a given star where planets are at the right distance from their host star
to potentially sustain liquid water at their surface provided a given atmospheric composition is present. In this context, potentially habitable planets relates to planets that loosely
resemble the Earth, made of iron and silicates and with a water content that can reach up to 50\% in mass, and orbiting in the habitable zone of their host star. 
 
\begin{figure}[tbp]
 \includegraphics[width=1.\linewidth]{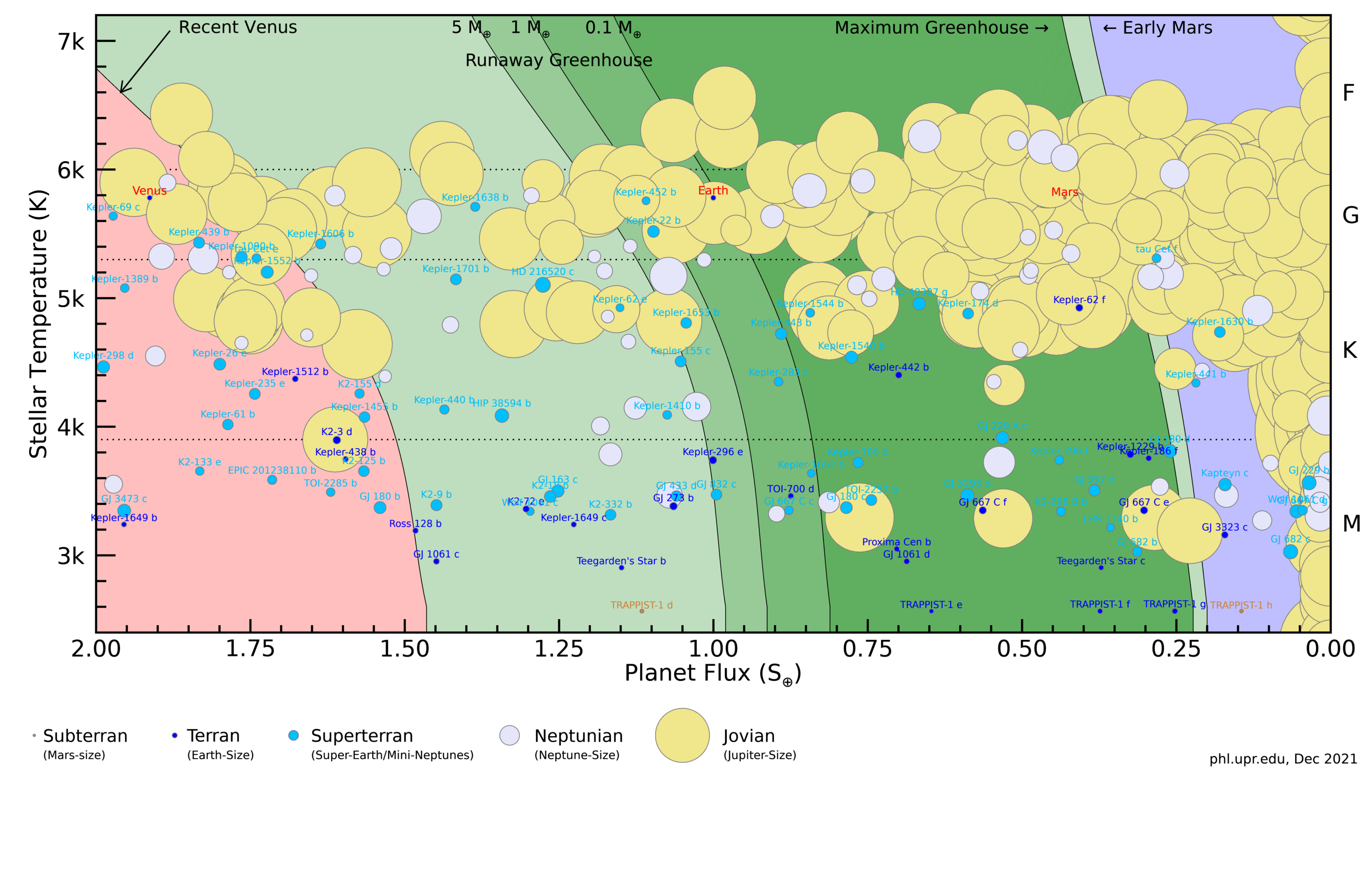}
 \caption{Distribution of the currently detected planets with the conservative and optimistic habitable zones as a function of stellar temperature of the host star and the flux received by the planet. The flux received is normalized to the Earth one. This plot is regularly updated and produced by the planetary habitability laboratory at the university of Puerto Rico www.phl.upr.edu.} 
 \label{fig2}
\end{figure}

This situation is illustrated in figure \ref{fig2} that shows a diagram elaborated by the planetary habitability laboratory at the university of Puerto Rico\cite{porto}. It displays the
distribution of planets in the habitable zone (see section 1.1) detected by the end of 2021 with the habitable zone boundaries discussed in the previous section. The planets are shown
as a function of their mass and are categorized following solar system counterparts. The continuum between subterran (Mars), terran (Earth-like), and superterran (Super-Earth-mini-Neptune)
are arbitrarily divided in two categories. Terran planets represent objects with radius and mass respectively less than 1.6R$_E$ and 3M$_E$. Superterran on the other hand are defined with a
radius and a mass greater than terran but lesser than respectively 2.5R$_E$ and 10M$_E$. These two broad categories can be associated to the dip in the size
distribution identified in the population of exoplanets detected by the Kepler mission\cite{fulton2017}. Analysis of the Kepler data shows a higher occurrence rate of planets at 1.3R$_E$
and 2.4R$_E$ with a valley at 1.8R$_E$ suggesting that there is two distinct populations of planets. Further analysis of the mass-radius relationship of planets currently detected
indicates that the first population matches Earth and Venus mean densities while the second population, with lower densities, have a higher fraction of volatiles and up to 50\% of
water\cite{zeng2019}. The latter objects do not exist in the solar system and are sometimes also refereed to as waterworlds\cite{mazevet2019}.

Figure \ref{fig2} shows that due to observational biases with current detection methods, no Earth-like planets have been detected so far in the conservative
habitable zone around F or G type stars. Earth-like planets detected in the conservative habitable zone are around K-stars (3500K<T<5000K) and  M-stars
(T<3500K). There are 13 objects at present with a highest occurrence around M-stars. This number reaches 20 when considering the optimistic habitable zone. It
can be extended to 55 when considering planets with a significant amount of water and categorized as superterran. This estimate of potentially habitable planets shows that so far,
about 0.2\% of the planets detected are labeled potentially habitable if we consider Earth-like planets and a conservative definition of the habitable zone. This number increases
to 1\% if one considers an optimistic definition of the habitable zone and planets stretching toward mini-neptunes.

The exact number of potentially habitable planets detected can vary depending on the precise criterium used. This is not an issue as it is commonly understood
that this definition relates to planets that are potentially habitable and among which only a subset may finally turn out to be trully habitable. It should thus
be understood as a way to identify targets for further investigation such as atmospheric characterization. Futhermore, habitability depends on other criteria than the mean density
and the flux received from the host star. At first, the stellar luminosity increases during the star lifetime. We saw above that within the habitable zone, a planet will be habitable
only if the atmosphere is suitable. For a planet to be habitable and remaing habitable during a significant fraction of time for life to develop requires that the atmospheric composition remains
suitable\cite{kasting1993,Catling2017}.

The build-up and long term evolution of the atmosphere of Earth-like planets is a process that is not fully constrained. It rests on an initial
budget of volatiles, primary degassing during the magmatic state and the long term geophysical activity, i.e. the evolution of the interior structure, that involves degassing of
various species and possibly various tectonic regimes. 
The situation is even more uncertain for the planets characterized as habitable at present as they are around M-dwarfs, tidally locked with a formation history and stellar
conditions that largely depart from what we know on Earth. Furthermore, when considering habitability from an Earth perspective, it appears relevant to also consider its formation in the
context of the history of the solar system\cite{raymond2017}, the low eccentricity of its orbit, its dynamical stability due to the moon, its rotation rate, interior processes
allowing plate tectonics to occur and subsequently its long term carbonate-silicate cycle to name a few particular features. Whether all these particularities are necessary to make a planet
habitable is an open question that the atmospheric characterization of exoplanets might help address in the coming decades\cite{apai2018}.  

\subsection{Upcoming prospects for observational characterization}
The atmospheric characterization of potentially habitable planets has been so far obtained for the TRAPPIST-1 system and for a very narrow IR band(1.1-1.7$\mu$m) using
the Hubble Space Telescope\cite{dewit2016,dewit2018}. The TRAPPIST-1 system consists of seven planets orbiting a M-dwarf star in a very compact configuration. The
densities of these planets are rather comparable and are slightly lower than Earth's mean density\cite{agol2021}. These planets are clearly terrestrial in nature
and three of them are in the habitable zone (d,e,f) \cite{gillon2017}. The featureless spectra obtained using the Hubble space telescope could only rule out that
the atmosphere of these planets were not made of primordial hydrogen. The limited spectral window available and the sensitivity of the instrument did not allow to identify
other volatiles such as H$_2$O or CO$_2$. The detection of these elements in the atmosphere of terrestrial planets, which can help validate the concept of habitability, and others,
such as O$_2$ or CH$_4$ related to biological activity, is just starting. 

The challenge ahead can be illustrated by considering the TRAPPIST-1 system in the era of the JWST. The distance of the TRAPPIST-1 system, 12pc, combined with the luminosity of the
late M-dwarf star provides a first opportunity to probe the atmosphere of Earth-size terrestrial planets with three of them in the habitable zone.
The characteristics of the planetary system suggest that transit spectroscopy could be performed for the seven planets detected so far if the instrument is performing as
expected.

A consensus is now emerging that the atmospheric properties could be obtained by using between 10 to 100 transits if these planets have a significant atmosphere\cite{gillon2020}.
CO$_2$ can be potentially detected using about 10 transits in a cloud-free atmosphere. This goes up to 20-30 transits if the atmosphere is cloudy. The detection
of H$_2$O is much more challenging if the planets resemble the Earth where the water is concentrated in the troposphere and is in small abundance at the altitude
probed by transit spectroscopy. If the atmosphere is different, detection of water vapor is estimated to require at least 60 transits but without giving indication whether liquid
water is present at the surface\cite{lincowski2019,lustig-yaeger2019,kopparapu2020}.
The JWST measurements can potentially give better constrains if the TRAPPIST-1 planets are entering a wet greenhouse runaway by detecting a strong massive oxygen atmosphere resulting
from ocean lost. Hence, the prospects of assessing the habitability of TRAPPIST-1 planets using JWST might principally concern the inference of the lack of greenhouse runaway. 

If we now turn to biosignatures, O$_2$, that could constitute a biosignature as it is associated to oxygenic photosynthesis, would require more than 100 transits. Detection of CH$_4$ could
be obtained with 10 transits for TRAPPIST-1e but its connection with biotic activities requires better contraints on the planetary context and to quantify the impact of biogical activity
on the planetary conditions to be conclusive\cite{Krissansen_Totton_2018,gillon2020,sauterey2020}.  With the TRAPPIST-1 system accessible 1/3 of the year and the number of possible
transits representing a fraction of that time, 65 transits possible for planet b and 5 for planet e, this suggests that the atmospheric characterization of the TRAPPIST-1 system may
require a significant fraction of the JWST lifetime. A few other systems presenting a comparable optimum configuration may be identified in the coming years but the identification of
exoplanet habitability and potential biosignatures is unlikely with this observatory.

The situation will improve with extremely large telescopes (ELTs) that will start operating toward the end of the decade. ELTs will increase the number of targets
accessible around M-dwarfs and maybe K type stars. It may provide more a precise characterization based not only on spectroscopy for transiting and reflected light measurements
but also by using direct imaging. It remains unlikely, however, that observations performed on a single or a few planets will provide compelling evidence of habitability or inhabitation
of Earth-like planets\cite{lehmer2020}. This is due to the difficulties at interpreting a single or even a few observations that necessarily rest on refined modeling.
Following \cite{bean2017} and others, we argue here that definite proof regarding habitability and inhabitation requires a statistical approach
to identify  trends in population of planets that will enable us to gain compelling evidence for the existence of worlds that are habitable and inhabited. This task might
be achieved using the next generation of space observatories based on direct imaging. This corresponds to mission concepts such as the Large UV/Optical/IR Surveyor
(LUVOIR)\cite{luvoir2019} or HABEX\cite{habex2020} that aimed at providing a statistically meaningful sample of objects to probe.
The design of the next generation space observatories is underway and there is currently a need to define the requirements these instruments should achieve to test unambiguously habitability
and inhabitation of extrasolar planets\cite{bixel2021}.  To contribute to this question, we consider in next section an innovative approach that adresses the question of habitability, the
validation of the habitable zone concept and the question of biosignatures by considering on the same footing the planetary context and the hypothetical biological activity.

\section{An integrated approach to assess habitability and inhabitation}
Following \cite{bean2017,bixel2020,bixel2021,lehmer2020}, we support here the idea that the identification of habitability and inhabitation should be achieved by identifying trends
in exoplanet populations rather than aiming at a precise characterization of a given planet. Considering the question of biosignatures, we propose that it should be approached
by directly coupling the planetary environment with the potential biological activity to account for the viability of potential metabolisms and to include their impact on
the atmospheric composition. This contrasts with more conventional approaches that aim at looking at the precise identification of a constituent related to
biological activities for a single planet. The case of oxygen that relates to oxygenic photosynthesis illustrates best such an approach\cite{kaltenegger2017}.
Secondly, we argue that it should be approached statistically as a trend in a population of habitable exoplanets. This formally reduces the possibilities of false positives
as it addresses the question of inhabitation of habitable planets as one of their global properties rather than the particularity of a single object.

To bring this approach on a quantitative basis, we developed a model that couples a 1-D photochemical-climate model, an interior model based on a time-dependent parameterization of the
carbonate-silicate cycle\cite{krissansen-totton-2017}, and an Earth-based ecosystem model to account for the biological activity\cite{sauterey2020}. This approach expands on the work of Karecha
et al. \cite{kharecha} who developed an atmosphere-ecosystem model
of the Archean Earth\cite{kharecha}. It also follows on the work of Lehmer et al.\cite{lehmer2020} who couples a 1-D climate model with a parameterization of the carbonate-silicate
cycle to assess trend in population of exoplanets to validate the habitable zone concept. It expands on this work by providing a quantitative modeling of the biological
activity that is consistent with the planetary environment. Furthermore, it provides an atmospheric composition that is in turn impacted by the biological activity. To illustrate
the features of the approach and demonstrate its usefulness for informing mission designs aiming at identifying habitability and biosignatures for exoplanet populations, we applied it to
the case of Enceladus\cite{affholder2021}, primitive Earth\cite{sauterey2020} and primitive Mars\cite{sauterey2022} and to the properties of Earth-like exoplanets around G-type
stars\cite{affholder2022}.

\subsection{Enceladus and the Cassini observations}
\begin{figure}[tbp]
 \includegraphics[width=1.\linewidth]{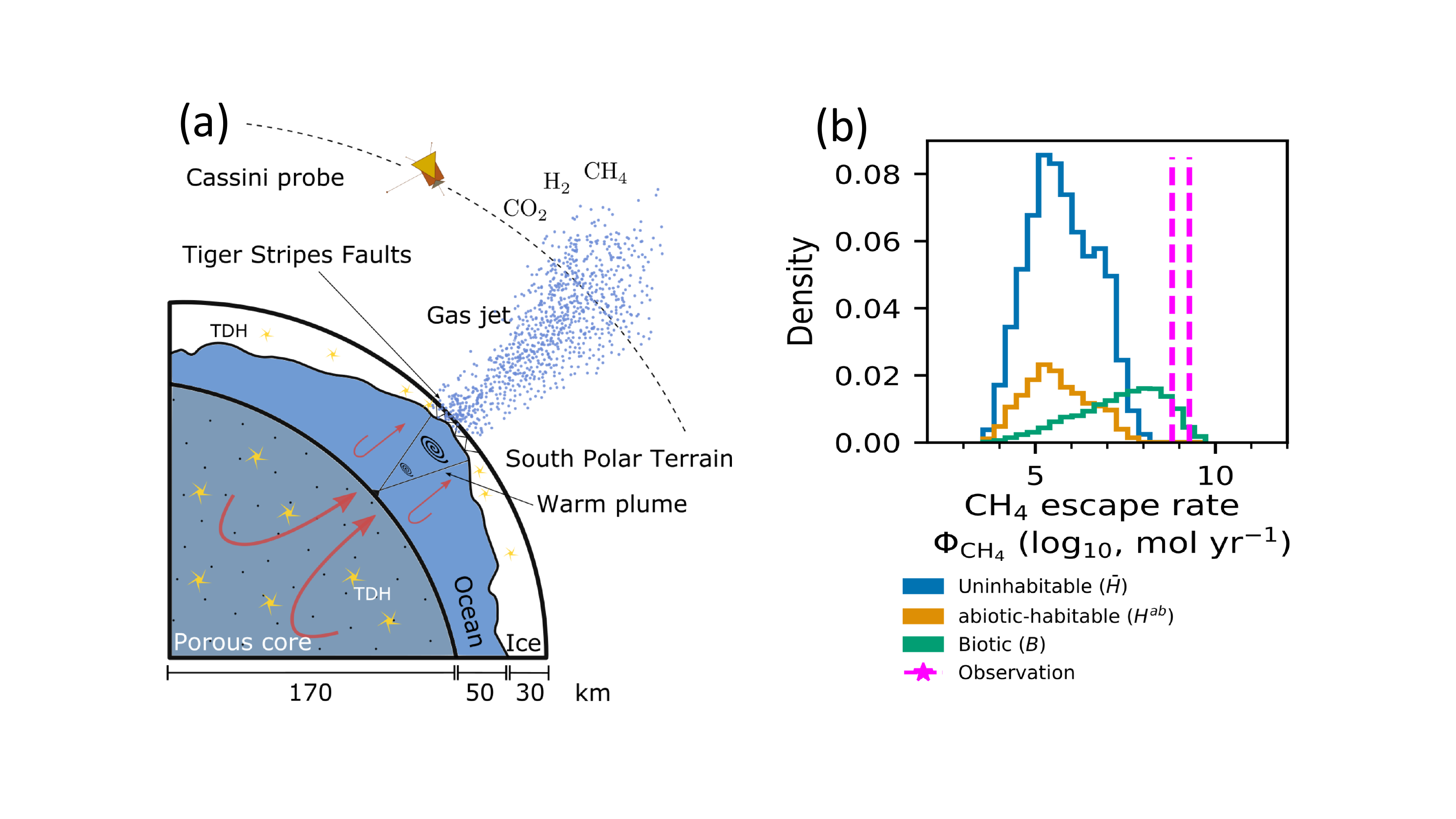}
 \caption{(a) schematic representation of the interior of Enceladus  (b)Density distribution of pseudo-data generated by model simulations for the inferred CH$_4$ flux (adapted from \cite{affholder2021}).} 
 \label{fig3}
\end{figure}
Multiple fly-bys of Enceladus by the Cassini mission confirmed that a global ocean is present\cite{spiker2019,thomas2016}. The mass spectrometer onboard the spacecraft
also revealed the presence of H$_2$ and CH$_4$ among other gases in plumes of gas and ice grains ejected into space at Enceladus's south pole. The detection of H$_2$ was interpreted as a
signature of hydrothermal activity\cite{waite2017}. On Earth, alkaline hydrothermal activity produces H$_2$ and CH$_4$ among other gases. In Affholder et al. \cite{affholder2021}, we
assume that hydrothermal circulation on Enceladus
  has its origin in a global ocean harboring hydrothermal vents and, in turn, that the composition of the plume provides information on the deep-sea vents. Figure \ref{fig3}-a
  shows the interior structure hypothesized for Enceladus with a core heated by tidal dissipation and ocean plumes produced by buoyancy and Coriolis
  forces\cite{choblet2017,goodman2012}. In this configuration, H$_2$ can be produced in Enceladus core by serpentinization\cite{choblet2017} while CH$_4$ might originate
  from the abiotic conversion of H$_2$ and CO$_2$, a primordial stock or organic molecules present in the core that are pyrolysed.

The biotic origin of CH$_4$ in Enceladus plume has also been largely discussed due to the possible similarities with the Earth's hydrothermal vents\cite{taubner2018}.
On Earth, these vents host chemoautotrophic single-celled organisms -called chemotrophs- that use thermodynamical disequilibrium generated by chemical gradients
rather than sunlight as energy sources.
Chemotrophic organisms use redox potential as energy source for biomass production. They convert H$_2$ and CO$_2$ into CH$_4$ and
could significantly contribute to the production of CH$_4$ detected in Enceladus plume. To test whether these organisms can sustain Enceladus's environment, grow and produce
 amounts of CH$_4$ compatible with the abundance in the plumes, an ecological model of methanogenesis constrained by the interior environment was developed.     
The ecological model describes the biomass dynamics of a generic H$_2$-based methanogenic metabolisms that represent archaea that have been found in Earth’s hydrothermal
vents. The model is first based on a dependence of the cell metabolism on environmental conditions through the thermodynamics and kinetics of the catabolic (energy-providing) reaction.
Catabolic reaction represents
the set of reactions a cell uses to maintain itself and grow. The cell population dynamics is described as
\begin{eqnarray}
  \frac{dN}{dt}&=&N(q_{ana}-d)\\
    \frac{dC_i}{dt}&=&F(C_i)+NB^*q_{cat}Y^{cat}_i
\end{eqnarray}
where $N$ is the surface density of the number of individuals in the population (kg$^{-1}$m$^{-2}$), $d$(s$^{-1}$) is the baseline death rate, $B^*$(mol$_{c_x}$) is the steady-state
internal cell biomass and $Y^{cat}_i$ the stoichiometry of the reactants and the products involved. $q_{ana}$ is the temperature dependent rate of net biomass build-up which
sets in turn the division rate.
This term depends on the Gibbs free energy required to produce 1 mol of biomass, the Gibbs free energy dissipated to maintain the cell
alive, and $q_{cat}$ the temperature-dependent rate at which the cell runs the catabolic reaction. $F(C_i)$  is the forcing term setting the concentration of species $i$,
$C_i$, given by the modelisation of the planetary environment. In the case of Enceladus, the H$_2$-based methanogenic population is assumed to be located at the core-ocean
interface and $F(C_i)$ represents the concentration calculated in a mixing layer between the hydrothermal fluid and the oceanic water.   

This model describes the dynamics of a population of H$_2$-based methanogens that convert H$_2$ coming from serpentinization with CO$_2$ present in the ocean to produce CH$_4$. The
population can reach different steady states depending on temperature and species concentration. This, in turn, sets the efficiency of H$_2$ consumption and conversion rate into
CH$_4$. It also provides a definition of habitability that corresponds, in this case, to the physical and chemical conditions allowing for a H$_2$-based methanogen population to grow and
reach a steady state. This is defined as the abiotic-habitable conditions. When the population reaches a steady state, it formally modifies the chemical and physical planetary conditions.
This is the biotic case that considers the planetary environment inhabited and modified by the population of H$_2$-based methanogens.

This geochemical and ecological modeling then feeds a statistical hypothesis testing framework by assuming that temperature and the composition of the hydrothermal fluid and ocean(estimated
close to 275K) are drawn from prior distributions that encompass uncertainties on their estimates.  This includes uncertainties for the serpentinization rate, composition of the ocean for
the species of interest and the temperature at the core-boundary. From these prior distributions, input parameters can be drawn for the dynamical model that defines a pseudo-data
distribution in the space of observables that corresponds, in the current case, to the flux of H$_2$ and CH$_4$ in the plume. Figure \ref{fig3} shows the distribution of model outputs
for the flux of CH$_4$ obtained by running 50,000 simulations. The model outputs are arranged in the three categories described above, unhabitable, abiotic habitable and biotic depending
on whether a methanogen population can grow and reach a steady state. When compared with the Cassini measurements, figure \ref{fig3} shows that the observed flux of CH$_4$ is not reproduced
by our abiotic model, but is well within predictions under the assumption of biotic methane production at the seafloor.

This result does not imply that there is an identification of an H$_2$-based methanogen population at the core-ocean boundary. It indicates that given the abiotic sources of
H$_2$ and CH$_4$ considered within the planetary model, the fluxes of CH$_4$ and H$_2$ observed are compatible with those expected when an H$_2$-based methanogen population is
present given the abiotic situation described. This assumes, first, that the probability of life to emerge for theses conditions is significant and, secondly, that no other abiotic
process can change significantly the production of H$_2$ and CH$_4$ modeled. While these two hypothesis can be challenged, and other can be added, this application illustrates
that the framework proposed allows us to do so within the bounds of a quantitative approach where the abiotic and biotic processes are treated on an equal footing and within an
integrated approach resting on a direct coupling between the two. We argue that this is a necessary step to identify habitable conditions and design experiments to identify
biological activity beyond Earth.  

\subsection{Primitive Earth and primitive Mars}

\begin{figure}[tbp]
 \includegraphics[width=1.\linewidth]{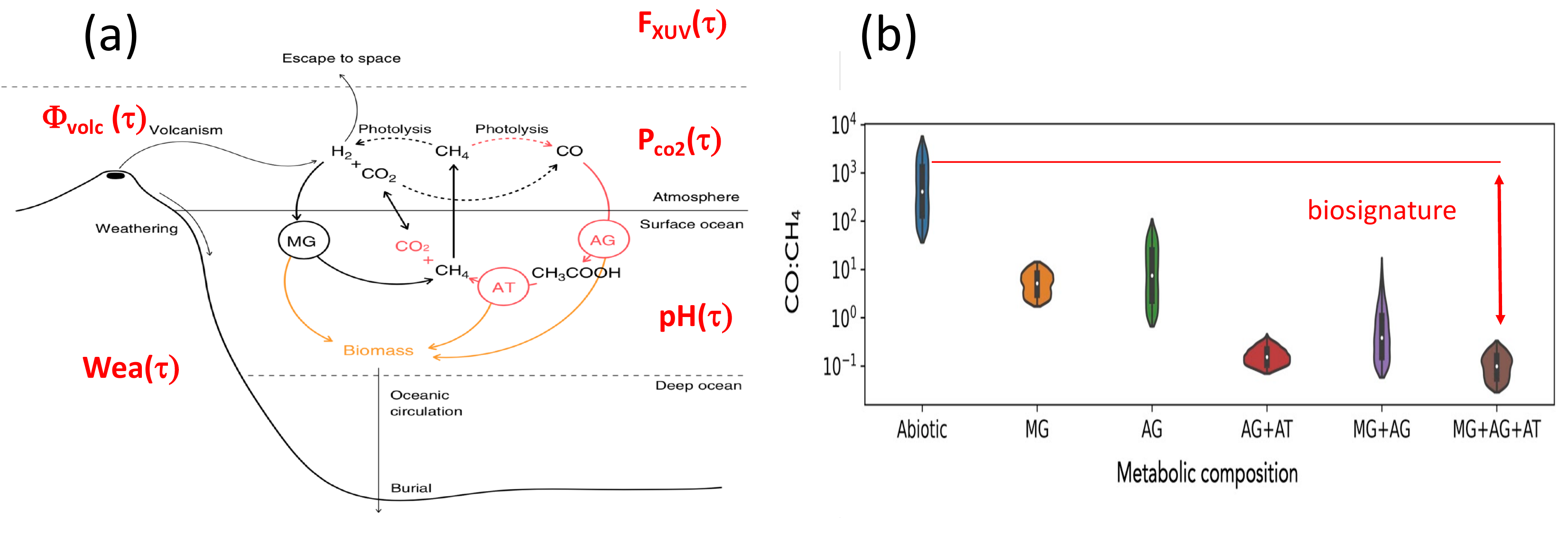}
 \caption{(a) Schematic representation of the Earth populated by a basic ecosystem involved in methanogenesis. The parameters that vary during the planet lifetime, $\tau$, are
   indicated in red. $\phi_{volc}(\tau)$ stand for the volcanic outgassing, Wea($\tau$) represents weathering, P$_{CO_2}(\tau)$ the CO$_2$ surface pressure, pH($\tau$) the ocean pH,
   and $F_{XUV}(\tau)$ the stellar flux. (b) Distributions of the CO:CH4 ratio for each plausible evolutionary scenario involving the basic metabolisms considered in the basic ecosystem.} 
 \label{fig4}
\end{figure}

In Sauterey {et al.}\cite{sauterey2020}, we apply the approach to quantify the effect of methanogenesis on the atmosphere of primitive Earth. Figure \ref{fig4}
shows that this extends the work for Enceladus described above in several ways. First, the ecosystem consists of not one but several populations of chemotrophic organisms that
can develop in the primitive ocean that covers part of the surface of the primitive Earth. Chemotrophs are expected to be the first organisms to develop on Earth, maybe several
hundreds millions of years before the apparition of anoxygenic photosynthesis. While this hypothesis is still debated, phylogenetic analysis combined with isotopic evidence
suggests that this primitive ecosystem may have consisted of H$_2$-based methanogens (MG), CO-based autotrophic acetogens (AG), and methanogenic acetotrophs (AT). To quantify
the effect of methanogenesis on anoxic Earth conditions, we considered these three metabolisms as well as anaerobic oxidation of methane (MT) as a basic ecosystem. 

The mean composition and temperature of the ocean and the atmosphere are obtained by using a climate and a photochemical models coupled to a simple temperature dependent
carbonate-silicate cycle\cite{krissansen-totton-2017}. The latter computes the evolution of atmospheric CO$_2$, dissolved inorganic carbon (CO$_2$, HCO$_3^-$ and CO$_3^{2-}$) and pH in
the ocean by taking into account outgassing from volcanos and mid-oceanic ridges, continental silicate weathering, dissolution of basalt in the seafloor, and oceanic chemistry, as sources
and sinks of CO$_2$. This coupled ecosystem-planetary model is solved by considering that the atmospheric temperature and composition and vertical structure reach equilibrium
instantly compared to the biological and geological time scales. Secondly, it considers that the dynamics of the ecosystem that occurs in days to years is much
faster than the geological time scale (10$^3$-10${^6}$ yrs). As previously, the ecosystem dynamics is obtained by considering fixed environmental conditions but given, in this case,
by a forcing term, $F(C_j)$, that results from ocean circulation and atmosphere-ocean exchanges and calculated using a stagnant boundary layer model\cite{kharecha}. The steady-state
 is obtained by considering  the biogenic fluxes between the surface ocean, the atmosphere, and the deep ocean iteratively within the planetary model.
 Once the global steady-state is obtained it represents the coupling between the biotic activity that modifies the planetary conditions that, in turn, feeds back into the efficiency
 of the biotic activity.

Applied to the case of primitive Earth, we first found that chemotrophic activity had a significant impact on the surface conditions of the anoxic Earth well before nonoxygenic
photosynthesis appeared and despite a rather small biomass production. The impact is the strongest on short time scale, 100Myrs, when a new metabolism is introduced
or removed from the atmosphere. This can lead to transient warming or ice ball events. On longer time scales, the carbonate-silicate cycle that includes the effect of the
biological activity through its dependence on temperature keeps the surface conditions habitable with temperature close to what is expected without biological activity but with a
different state for the atmospheric composition. The imprint of the individual metabolism on the atmospheric composition and surface temperature is clear but converges to a common
state when the ecosystem becomes more complex and several metabolisms are included. Figure \ref{fig4} illustrates this situation when considering the CO/CH$_4$ ratio of these two species
in the atmosphere. We see that the atmospheric composition changes as soon as a metabolism is included but the footprint of the individual metabolism washes out when the ecosystem becomes
more complex to leave a clear signature between abiotic and biotic biospheres.           

This study shows that the atmosphere of the anoxic Earth was modified as soon as life emerged. The direct coupling between the ecosystem dynamics and the planetary conditions further
allowed us to quantify how methanogenesis impacted planetary conditions, showing that Earth remained habitable on geological time scale. A similar model applied to the case of primitive Mars
showed that methanogenesis likely drove Mars out of habitability if these primitive metabolisms appeared\cite{sauterey2022}. This quantitative modeling suggests a very different outcome
between the Earth and Mars upon the potential apparition of methanogenesis, stressing the need for an integrated approach to address habitability and
inhabitation on the same footing. This also shows that habitability is, to some extent, determined by inhabitation. The common steady state obtained for primitive Earth in the
biotic case, distinct from the abiotic one, and reached as the ecosystem becomes more complex, suggests that a biosignature can be defined. As shown in figure \ref{fig4}, this biosignature
is rather independent of the particular characteristic of the metabolisms involved suggesting
that this could be used as a biosignature of methane cycling biospheres for future surveys of habitable planets.

\subsection{Earth-like population around G-type stars}

We now consider the case of Earth-like planets around G-type stars to illustrate how an integrated ecosystem-planetary model can be used to inform the design of future space telescopes
aiming at characterizing habitability and inhabitation of extrasolar planets.  We saw in the the first section that the definition of the conservative habitable zone rests on climate
modeling that embodies the greenhouse effects of H$_2$O and CO$_2$ present in the atmosphere. While not explicitly included, it assumes that a carbonate-silicate cycle that
regulates the amount of atmospheric CO$_2$ needed to keep the planet habitable exists as the stellar luminosity increases during the host star lifetime \cite{kasting1993,Catling2017}.
By demonstrating that a population of Earth-like planets follows the general concentration
of CO$_2$ anticipated would first validate the concept of habitable zone. It would secondly indicate that these planets may have liquid water on their surface as the carbonate-silicate
cycle as we experience it on Earth depends on having liquid water at the surface. Incidentally, it would also confirm how the Earth stayed habitable for 4 billions years (4Gyrs).
This approach  was suggested by Lehmer\cite{lehmer2020} and implemented in a very simplified way to validate the habitable zone concept.

The integrated ecosystem-planetary model we developed for the case of primitive Earth allows us to go a step further by not only considering the evolution of the CO$_2$ concentration as
a function of time but by also following the complete evolution of the biosphere coupled to the carbonate-silicate cycle. The results shown in the previous section indicate that
this may be necessary as habitability and inhabitation are strongly coupled with inhabitation modifying habitability. This approach also provides an appealing way to identify inhabited
planets as biosignatures, such as the CO/CH$_4$ ratio mentioned above, can be used and interpreted by fully accounting for the planetary context.

\begin{figure}[tbp]
 \includegraphics[width=1.\linewidth]{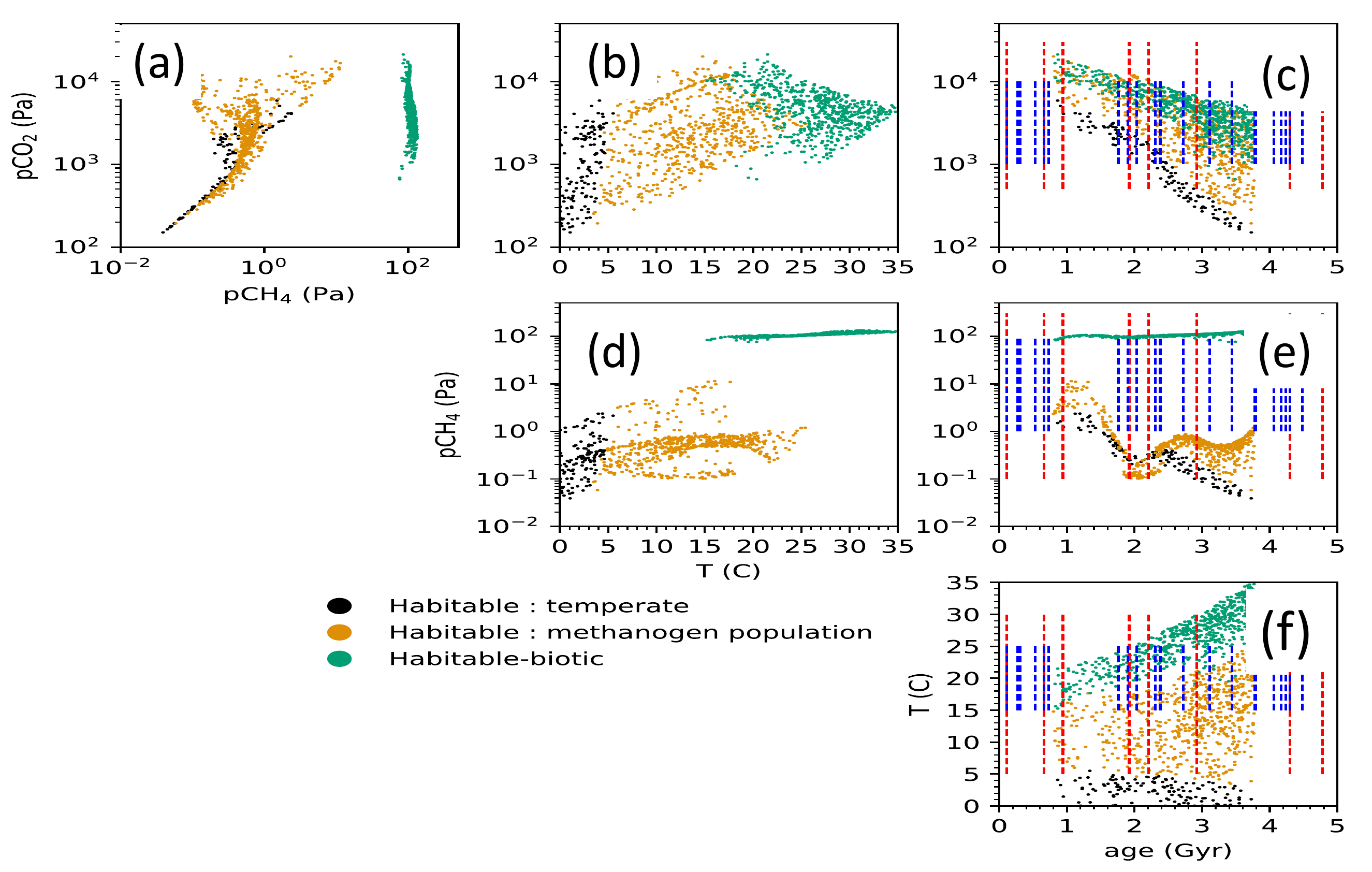}
 \caption{Results of simulated steady state atmospheres displayed as a function of the CO$_2$ surface pressure, P$_{CO_2}$, CH$_4$ surface pressure P$_{CH4}$, surface temperature, T, and time, t. The red and bleu lines indicate Earth-like planets with densities respectively close to the Earth (0.9<$\rho$<1.1) and defined as terran (see section 1.1) anticipated as detectable using an instrument configuration corresponding to LUVOIR-A. }
 \label{fig5}
\end{figure}

Figure \ref{fig5} shows preliminary results considering a primitive ecosystem consisting of a single and the same generic methanogen as the one used for the case of Enceladus. Each point
shown in figure \ref{fig5} corresponds to a complete simulation involving a similar calculation as performed for the case of primitive Earth, coupling photochemistry, climate, the
carbonate-silicate cycle and ecosystem modelings and taking into account the variation of volcanic outgassing, weathering and solar luminosity over 4.5 Gyrs as indicated in figure \ref{fig4}.
We note that several technical adjustments are needed to achieve this generalization of the primitive Earth model discussed above over 4.5Gyrs\cite{affholder2022}. Most of them remove
simplifying assumptions for the atmosphere-ocean-ecosystem model that are no longer valid when considering the large parameter space explored over such a long time scale, $\tau$.

In figure \ref{fig5}, the simulations are displayed in three categories. The simulations leading to a surface temperature above 273K are tagged as habitable-temperate as surface temperature
is the only criteria used. Following what was done for Enceladus, simulations where a methanogen population can reach a steady state are tagged as habitable for the methanogen population.
This represents a fraction of the habitable-temperate set as in addition to temperature, the ocean composition enters in defining the viable conditions for methanogens. Finally, the third
set of simulations, habitable-biotic, represents the conditions where the ecosystem activity feeds back into the atmospheric composition. It represents the steady state reached by the
ecosystem-planetary system at a given time, $\tau$, given by the stellar luminosity, volcanic outgassing and weathering. The latter two are obtained by the parameterization of the Earth
carbonate-silicate cycle\cite{krissansen-totton-2017} while the variation of stellar luminosity and spectrum follow the parameterization of Claire {\sl et al.}\cite{claire2012}. We note
that each simulation point is calculated independently.

Figure \ref{fig5} shows the variation  of the surface temperature and of the CO$_2$ and CH$_4$ surface pressures as a function of time and corresponding to the three sets defined above. We see in
figure \ref{fig5}-c that the amount of CO$_2$ decreases as time evolves and the stellar luminosity increases to maintain habitable conditions. We note that the sun luminosity increases
by 25\% over 4 Gyrs in these simulation sets. This trend in the atmospheric concentration of CO$_2$ as a function of time represents the effect of the carbonate-silicate cycle.
This variation also translates into a varying concentration of CO$_2$ across the habitable zone. 
  Figure \ref{fig5}-e shows that the methanogenic population brings an almost constant increase of CH$_4$ in the atmosphere. This biogenic activity is associated to an increase in surface
  temperature that can be seen in figures \ref{fig5}-d-f due to the greenhouse effect of CH$_4$. Inspection of figure \ref{fig5} indicates that clear trends exist for the evolution of the
  atmospheric concentration of CO$_2$ and CH$_4$ over time that may allow us to identify habitability and inhabitation in population of exoplanets. We note that habitability and inhabitation can
  be accessed independently by considering respectively the habitable temperate and habitable-biotic sets.

  If we take the history of life on Earth and its impact on the atmospheric composition of the planet as a template, biogenic CH$_4$ due to chemotrophic activity and non-oxygenic photosynthesis
  was dominant for more than 1 Gyrs. It was followed by a rise in oxygen related to the apparition of oxygenic photosynthesis in two steps, with detectable amount in the atmosphere only
  800Myrs ago. The trends identified in this preliminary calculation correspond to planets that exhibit a carbonate-silicate cycle similar to the one experienced by the Earth and a
  methanogenic population pertaining over 4 Gyrs without the emergence of non-oxygenic and oxygenic photosynthesis. These preliminary simulations can be expanded to explore more throughly
  the parameter space when chemotrophic metabolisms are at play. Photosynthetic metabolisms could be formally included in the approach but require substantial modifications of
  the ecosystem formalism. The simulations show nevertheless that priors can be built to be used in a Bayesian analysis scheme using a population of potentially habitable planets.     

  Figures \ref{fig5}-c-e-f illustrate this approach by displaying the exoplanet yield for Earth-like planets around G-type stars expected from the next generation of space telescopes.
  The sets of exoplanets shown are obtained by using the Bioverse scheme\cite{bixel2020,bixel2021}  and considering a configuration similar to the LUVOIR-A mission concept\cite{luvoir2019}. The
  Bioverse scheme simulates planetary systems for host stars in the solar neighborhood based on the {\sl Kepler} statistics\cite{Kopparapu_2018,Pascucci_2019}. The first set displayed in
  figure \ref{fig5} corresponds to planets anticipated in the habitable zone, with densities between 0.9 and 1.1 the Earth one, and potentially detectable using direct imaging. This set corresponds
  to planets very close to the Earth and could be associated to Earth twins. At first glance, these planets would have a higher probability of displaying a carbonate-silicate cycle similar to the
  Earth. The second set considers a less restrictive definition and corresponding to the one given in section 1.2 to define terran planets. We see in figure \ref{fig5} that considering planets
  with densities very close to the Earth, the detection of only eight planets are anticipated with this particular telescope configuration. This number significantly increases when considering
  planets with a less restrictive density criteria. These precise numbers may be refined in the near future as more surveys come along but it gives a good estimate of the number of candidates
  that can be expected. Considering that the space observatory has the ability to detect these levels of atmospheric CO$_2$ and CH$_4$,
  figure \ref{fig5} shows that the number of planets anticipated with a configuration corresponding to LUVOIR-A will likely be too low to identify a trend in the variation of CO$_2$
  as a function of time for planets with densities very close to the Earth one. The configuration of the LUVOIR-A mission concept may, in contrast, be sufficient to identify methanogenesis
  if all the candidates
  display a constant and similar level of CH$_4$ or by considering planets defined as terran with physical properties departing from Earth twins. 

  This example illustrates how these studies can be used to design instruments and devise observational strategies aiming at identifying trends in populations of exoplanets. For example,
  the carbonate-silicate cycle as parameterized for the Earth depends on the ability of a planet to maintain liquid water at its surface. Detecting trend in CO$_2$ as diplayed in figure 4 instead
  of water in the atmosphere are two different approaches that could formally be used to identify habitable planets. This leads to two different designs for the space observatory. Furthermore, when
  considering the atmospheric concentration of CO$_2$ of a population of planets in the habitable zone of G-type star and with densities close to the Earth, we see that the statistical set
  expected is too limited when considering the configuration of the LUVOIR-A mission concept.  This can be overcome by considering a radically different approach for the design of the space
  observatory\cite{apai2019}. Conversely, the planet sample can be expanded by considering a less restrictive definition of Earth-like planets to include a larger variation in mass, radius and
  density as displayed in figure 4. This, however, requires that the current modelisation folds in the dependance of the carbonate-silicate cycle on these parameters to identify trends.
  Alternatively, it can also be approached by using a larger set of stars involving different types and considering luminosity instead of time. Considering planets with a carbonate-silicate cycle
  driven by different geophysical regimes such as plaque tectonics like on Earth or a stagnant lid regime like on Venus may also devise a better observational strategy than considering a
  log-linear relationship between luminosity and CO$_2$ concentration as suggested in \cite{lehmer2020}. Overall, we see that the ecosystem-planetary model developed used in combination
  with the Bioverse simulations provides a powerful tool to address these questions quantitatively and progress toward the identification of habitable and inhabited planets.

\section{Summary}

The characterization of habitable planets and the possible detection of biotic activity is a major challenge for the coming decades. It will likely require the development of a new generation
of space telescope optimized to identify trends in population of exoplanets rather than focused on the precise characterization of a few objects. The design as well as the future analysis of
these experiments require a new generation of models that integrate the coupling between the potential biotic activity, the atmospheric and interior evolution of terrestrial planets that
expand beyond the Earth size and density. We develop such an approach that couples an Earth-based ecosystem and planetary modeling. We demonstrated that is is applicable to identify biotic activity
and quantify its effect on habitability for Enceladus, primitive Earth and primitive Mars, as well as for a population of Earth-like planets around G-type stars. The latter represents a
preliminary analysis that can be expanded in several directions to asses trends in population of exoplanets that can potentially bring the concept of habitable planets to
observational confirmation.   

\section{acknowledgment}

The results reported herein benefited from collaborations and/or information exchange within the program “Alien Earths” (supported by the National Aeronautics and Space Administration under
Agreement No. 80NSSC21K0593) for NASA’s Nexus for Exoplanet System Science (NExSS) research coordination network sponsored by NASA’s Science Mission Directorate.

\bibliographystyle{crunsrt}

\nocite{*}

\bibliography{bib.academie}

\end{document}